\documentclass[preprint,eqsecnum,aps,showpacs,floatfix]{revtex4}
\usepackage{dcolumn} \usepackage{graphicx} \usepackage{amsmath}
\usepackage{amssymb} \usepackage{epsfig} \usepackage{float}
\usepackage{latexsym} \usepackage{rotating}
\begin{document}
\title{Confinement of test particles in warped spacetimes}

\author{
Suman Ghosh\footnote{Electronic address: {\em suman@cts.iitkgp.ernet.in}}${}^{}$}
\affiliation{Department of Physics and Meteorology and Centre for Theoretical Studies  \\
Indian Institute of Technology, Kharagpur 721 302, India}
\author{
Hemwati Nandan\footnote{Electronic address: {\em hnandan@cts.iitkgp.ernet.in}}${}^{}$}
\affiliation{Centre for Theoretical Studies  \\
Indian Institute of Technology, Kharagpur 721 302, India}
\author{
Sayan Kar\footnote{Electronic address: {\em sayan@cts.iitkgp.ernet.in}}${}^{}$}
\affiliation{Department of Physics and Meteorology and Centre for Theoretical Studies  \\
Indian Institute of Technology, Kharagpur 721 302, India}

\begin{abstract}
We investigate test particle trajectories in warped spacetimes with a 
thick brane warp factor, a cosmological on--brane line element and a 
time dependent extra dimension. 
The geodesic equations are reduced to a first order autonomous dynamical 
system. Using analytical methods, we arrive at some useful general conclusions 
regarding possible trajectories. Oscillatory motion, suggesting
confinement about the location
of the thick brane, arises for a growing warp factor. On the other hand,
we find runaway trajectories (exponential-like) for a decaying warp factor.   
Variations of the extra dimensional scale factor yield  certain 
quantitative differences. Results obtained from explicit numerical
evaluations match well with the qualitative conclusions obtained
from the dynamical systems analysis.

\end{abstract}

\maketitle

\section{Introduction}
Much of the initial interest and subsequent research on the hypothesis that we
live on an embedded, timelike submanifold (the brane) 
of a higher dimensional ($D>4$) 
Lorentzian spacetime (warped or unwarped) 
was focused on the assumption that the scale of the
extra dimension is independent of the coordinates. The original
work of Randall and Sundrum (RS) \cite{Rs1,Rs2} on warped braneworlds,
published a decade ago, did refer to the idea
of the scale of the extra dimension being
spacetime dependent, while addressing the issue of stability, in a two--brane
scenario. In such a model, it is necessary
to stabilise the inter-brane distance (modulus) to a fixed value, 
in order to avoid a collapse of one brane onto another. 
A spacetime dependent bulk scalar field, was used to achieve the
desired stability of the two--brane system, through the
Goldberger--Wise mechanism \cite{Gw}. A good
deal has since then been done with the assumption of a spacetime
dependent radion field, including its effects, through various types
of couplings, on particle phenomenology \cite{radionpp1,radionpp2}
associated with the two--brane RS model.

In a single brane scenario or from a purely higher dimensional
bulk perspective, the space-time dependence of the metric
function(s) associated with the extra dimensional coordinate(s)
basically imply that the scale of the extra dimension 
depends on the on--brane or four dimensional spacetime coordinates.
To visualise this, it is easiest to go back to the standard
Kaluza--Klein (KK) universe models with different scale factors
associated with the evolution of each set of non-compact/compact dimensions
(usual or extra). The difference between today's warped braneworlds
is the warped geometry and also the non--compact extra dimension(s).
We shall be concerned with single brane scenarios in this article. 

Since the early days of General Relativity, one possible way of understanding
the nature of the gravitational field has been to 
study geodesics and geodesic deviation \cite{Wald}-\cite{hartle} 
in the given background geometry. While the spacetime variation of the
connection reflects on the trajectories, the effects of curvature
variations clearly control geodesic deviation. This is true in any dimension 
and in any metric theory of gravity. Usually, generic statements 
(eg. existence of orbits of different types, oscillatory/exponential
behaviour in trajectories, focusing and defocusing etc.) and their relation
to variations in the metric functions, are
difficult to extract, though the primary goal in such studies
(on geodesics) must always
be so.     

Over the years, several authors \cite{geodesics1,geodesics2, Rk} have
investigated the behaviour of geodesics in 
background geometries of various higher dimensional 
theories of gravity. 
Recently, in \cite{seahra}, it was shown briefly, that in warped
product spacetimes it is possible to have classical confinement
of test particles.
In our work here, we confirm and add to the results in \cite{seahra}, through
detailed analysis, both numerical and analytical.  

Let us assume a bulk line element of the form,
\begin{equation}   
ds^2=e^{2 f(\sigma)} \left [-dt^2 + a^2(t)\,  d{\bf X}^2 \right ] + b^2(t)\, d\sigma^2  \label{eq:metric},
\end{equation}
where  $d{\bf X}^2 = dx^2+dy^2+dz^2$.  Here, the function $b(t)$ represents the scale of the
extra dimension while the $a(t)$ and $e^{2f(\sigma)}$ are the usual cosmological scale
and warp factors respectively. 
In what follows, we shall largely be concerned with the effect of the 
geometric properties of the bulk spacetime on the trajectories
that can exist in them. We intend to delineate how features
change when we vary the nature of each of the three functions $a(t)$, $b(t)$ 
and $f(\sigma)$. The geodesic equations cannot be solved analytically 
(modulo a few simplistic cases) in a spacetime as complicated as represented by (\ref{eq:metric}). Thus, we make use of the dynamical systems approach as well
as numerical methods, through which 
we are able to extract some useful information regarding the behaviour of 
trajectories. For numerical computations of the 
differential equations involved in this work, we have used standard 
numerical codes, with appropriate initial conditions on the variables associated with null and timelike 
geodesic evolution.\\
The paper is organised as follows. In the next Section, we elaborate on the spacetime geometry and discuss our choices for the functions $a(t)$, $b(t)$ and $f(\sigma)$ which we use henceforth. Then, in the subsequent Sections, we discuss the nature of geodesics in detail for our specific choices of the functions. 
Finally, we make a comparison of our results with those in certain limiting 
scenarios and conclude with a summary of the results obtained.
 
\section{The background spacetime geometries}


We consider the metric given by equation (\ref{eq:metric}) rewritten in conformal time $(\eta)$, as follows,
\begin{equation}
ds^2 = e^{2f(\sigma)} a^2(\eta) [- d\eta^2 + d{\bf X}^2] + b^2(\eta) d\sigma^2 .  \label{eq:cmetric}
\end{equation}
To arrive at concrete results in the following Sections, 
we need to choose the functional forms of the warp factor, cosmological and extra dimensional scale factors. 
For the warp factor $e^{2f}$ we choose,
\begin{equation}
f(\sigma) = \left\{\begin{array}{ll} -log\,(\cosh\, k\sigma) \hspace{.5 cm}\rightarrow \hspace{.5 cm}\mbox{represents a decaying warp factor,} 
                                  \\ \hspace{.3cm}log\,(\cosh\, k\sigma)\hspace{.5cm}\rightarrow \hspace{.5 cm}\mbox{represents a growing warp factor,} 
                   \end{array} \right. \label{eq:warpfactors}
\end{equation} 
which correspond to the well--known thick brane models \cite{Rk1, Kanti}.
In such models, the brane is dynamically generated as a scalar field
domain wall (soliton) in the bulk. Note the warp factor in such models
is a smooth function of the extra dimension, unlike the RS case where
we have $f(\sigma)= - k\vert \sigma \vert$ (i.e. a function with
a derivative jump--such branes are called thin branes). 
We choose to work with thick branes mainly 
to avoid the jumps and delta functions which will appear 
in the connection and curvature for thin branes.    

What do we choose for the scale factors?
Two different combinations of $a(\eta)$ and $ b(\eta)$, chosen as models 
to represent different kinds of time evolution, 
are given below in terms  of the cosmological time ``$t$'',
\begin{equation}
\{a(t), b(t)\} = (i)\hspace{.2cm}\{a_0 t^{\nu_1}, b_0 + b_1 t^{-\nu_2}\},  \hspace{.4cm} \mbox{and}\hspace{.4cm} (ii) \hspace{.2cm}\{a_0' e^{H(t-t_0)}, b_0' + b_1'e^{-\beta H(t-t_0)}\}, \label{eq:cosmo_sf}
\end{equation}
where we take $\nu_1$ to span over an open interval $(0,1)$, so that 
in $(i)$ $a(t)$ represents an expanding but decelerating on--brane line 
element (which is radiative for $\nu_1 = \frac{1}{2}$), and in $(ii)$ 
we take $H$ to be positive, representing an accelerating de--Sitter on--brane 
line element. $b_0$, $b_1$, $\nu_2$, $b_0'$, $b_1'$ and $\beta$ are positive 
so that we have a decaying extra dimension which stabilizes to a finite value 
as $t \rightarrow \infty$. The type of metric functions we consider here 
may be obtained as
analytic solutions of the five dimensional Einstein equations with 
the corresponding Einstein tensors providing matter energy-momentum 
profiles through the Einstein field equations. 
As shown in \cite{sols},
for similar metric functions, the corresponding matter stress-energy
can indeed satisfy the Weak Energy Condition. 

In the first case, the conformal time $\eta = \int{\frac{dt}{a_0 t^{\nu_1}}}$. 
We choose to work with an initial condition such that at $t = t_0$, 
$\eta = \eta _0 = 1$ which, in turn, implies 
$a_0 = \frac{t_0^{1-\nu_1}}{1-\nu_1}$ and $\eta = \left(\frac{t}{t_0}\right)^{1-\nu_1}$ where the domain of $\eta$ is $1 < \eta < \infty$.
We also assume that evolution of our universe begins at 
$t = t_0$ (or $\eta = 1$) when the scales of ordinary and extra space 
dimensions were same (i.e. $a(t_0/\eta _0) = b(t_0/\eta _0))$. In terms of 
conformal time, we are led to,
\begin{equation}
a(\eta) = \frac{t_0}{1-\nu_1} \eta^{\frac{\nu_1}{1-\nu_1}} \hspace{.3cm} \mbox{and}\hspace{.3cm} b(\eta) = \frac{t_0}{1-\nu_1} - \frac{b_1}{t_0^{\nu_2}} \left(1 - \eta^{\frac{-\nu_2}{1-\nu_1}}\right). \label{eq:confo_psf}
\end{equation}
The stabilization condition further implies,
\begin{equation}
b_0 =  \frac{t_0}{1-\nu_1} - \frac{b_1}{t_0^{\nu_2}} > 0 \hspace{.3cm} \mbox{or}\hspace{.3cm} b_1 < \frac{t_0^{1+\nu_2}}{1-\nu_1}. 
\end{equation}
In order to reduce the number of free parameters, we consider  
$b_1 = t_0^{1+\nu_2}$  (which is, in fact, 
the maximum possible value for $b_1$) as $\nu_1$ can only be close to zero (but not equal to zero). Therefore we have,
\begin{equation}
b(\eta) = t_0 \left(\frac{\nu_1}{1 - \nu_1} + \eta^{\frac{-\nu_2}{1-\nu_1}}\right). \label{eq:confo_pesf}
\end{equation}
Using a similar prescription for the de--Sitter model,  we obtain,
\begin{equation}
a(\eta) =  \frac{1}{H}\frac{1}{(1-\eta)}  \hspace{.3cm} \mbox{and}\hspace{.3cm} b(\eta) =  \frac{1}{H}\Big[1 - b_1' H \big\{1 - (1 - \eta)^\beta\big\}\Big], \label{eq:confo_dsf}
\end{equation}
where $b_1' H < 1$ and $0 < \eta < 1$. Let us consider, $\nu_1 = \nu_2 = \frac{1}{2}$, $t_0 =1$, $H = 1$, $b_1' = \frac{1}{2}$ and $\beta = 1$ which leads to the following two different combinations of the scale factors, 

(A)\, \,   $a(\eta) = 2\eta,\, \, \, \, \, \,  b(\eta) = 1 + \frac {1}{\eta}$, 

(B) \, \, $a(\eta) = \frac {1} {1 - \eta},\, \, \,  b(\eta) = 1 - \frac{\eta}{2}$.

\begin{figure}[!ht]
\fbox{\includegraphics[width = 2 in]{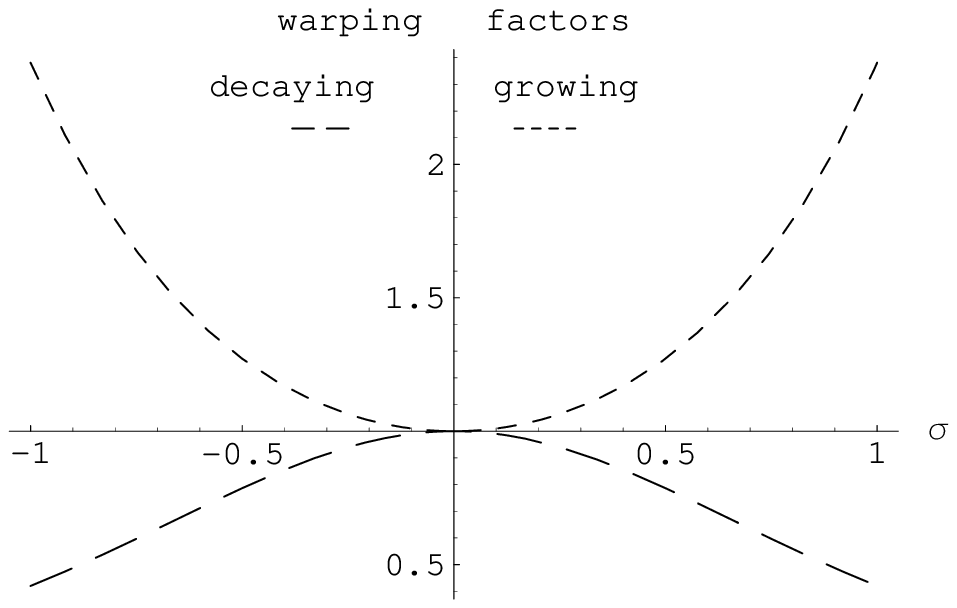}
\includegraphics[width = 2 in]{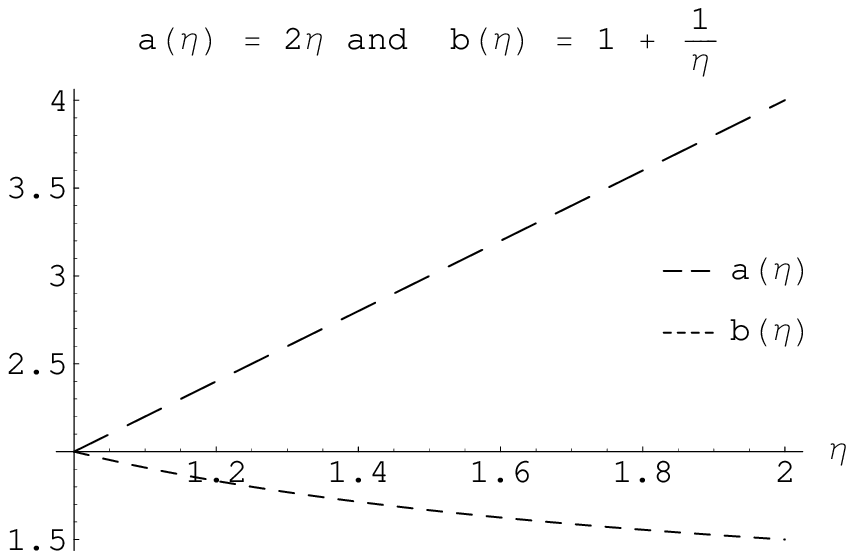}
\includegraphics[width = 2 in]{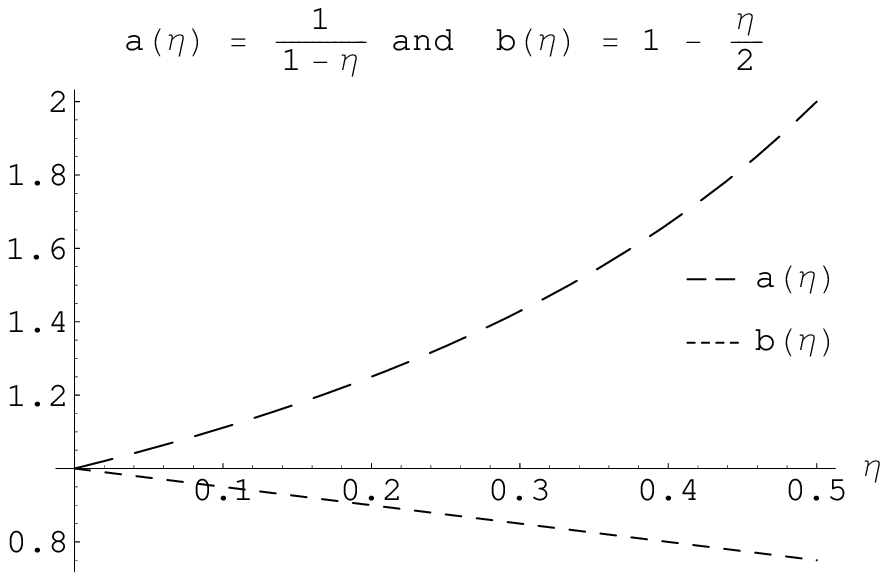}}
\caption{Variations of warp factors and the scale factors for set(A) and set(B) respectively.} \label{fig:models}
\end{figure}

\noindent These combinations of scale factors will henceforth be abbreviated as set (A) and set (B). Fig. \ref{fig:models} gives a pictorial representation of the warp factors (decaying and growing)  and the scale factors corresponding to set (A) and set (B).

\section{Geodesics}
 The general form of the constraint 
(i.e. $g_{AB} u^{A} u^{B} = -\epsilon)$ for null and timelike geodesics 
corresponding to the metric (\ref{eq:cmetric}) is given below,
\begin{equation}
e^{2f(\sigma)} a^2(\eta)\,  [- \dot \eta^2 + \dot {\bf X}^2 ]\, + b(\eta)^2 \, \dot \sigma^2 + \epsilon = 0, \label{eq:conmetric}
\end{equation}
where $\epsilon = 1$ and $0$ denote the cases corresponding to the 
timelike and null geodesics respectively and a dot here represents the differentiation with respect to the affine parameter $\lambda$. As $x_i$  (such that $x_1 = x$, $x_2=y$, and $x_3=z$) is cyclic, an obvious conclusion for the
metric (\ref{eq:cmetric}) is,  
\begin{equation}
\dot x_i = \frac{D_i e^{-2f}}{a^2} \label{eq:xcyclic},
\end{equation}
where  $D_i$'s are integration constants. The effective geodesic potentials $V_\sigma$ and $V_x$ are then calculated as, 
\begin{equation}
 V_\sigma(\lambda) = \frac{\epsilon +  a^2(\eta)  e^{2f(\sigma)}(-\dot \eta^2 + 3 \dot x^2)}{2 b^2},\label{eq:vs}
\end{equation}
\begin{equation}
 V_x(\lambda) = \frac{1}{6} \, \left [{\dot \eta^2 + \frac{\epsilon + b^2 \dot \sigma^2}{e^{2f(\sigma)} a^2(\eta)} } \right] = -\frac{D_i^2\, e^{-4 f(\sigma)}}{2a^4(\eta)}, \label{eq:vx}
\end{equation}
where we have considered $\dot x = \dot y =\dot z$, which will be used throughout hereafter. In order to find $V_\sigma(\sigma)$ (or $V_x(x)$), we shall make use of solutions for $x(\lambda)$, $\eta(\lambda)$ and $\sigma(\lambda)$ in the right hand side of equation (\ref{eq:vs}) (or equation (\ref{eq:vx})) and, subsequently, make a parametric plot of $V_\sigma(\lambda)$ versus $\sigma(\lambda)$ (or $V_x(\lambda)$ versus $x(\lambda)$) to arrive at a pictorial representation of $V_\sigma(\sigma)$ (or $V_x(x)$). It may be noted that, we have absorbed the factor ``$\epsilon$'' in the effective geodesic potential. Therefore, for both timelike and null geodesics, we are essentially looking at test particles with zero ``total energy'' (from a mechanics point of view).

\subsection{Dynamical systems analysis}

The geodesic equations corresponding to the metric (\ref{eq:cmetric}) derived for general warp and scale factors turn out to be,
\begin{equation}
\ddot \eta + \frac{a'(\eta)}{a(\eta)} \,\, [\, \dot \eta^2 + \textstyle{\sum_ i} (\dot x_i)^2 \, ] +  e^{-2 \, f(\sigma)} \,\,  \frac{b(\eta)\, b'(\eta)}{a^2(\eta)}\, \, \dot \sigma^2 + 2 f'(\sigma) \, \dot \eta \, \, \dot \sigma = 0,  \label{eq:etaeqn}
\end{equation} 
\begin{equation}
\ddot x_i + \frac{2 a'(\eta)}{a(\eta)} \, \dot \eta \, \dot x_i + 2 f'(\sigma) \, \dot x_i \, \, \dot \sigma = 0, \label{eq:xieqn}
\end{equation}
\begin{equation}
\ddot \sigma + \frac{2 b'(\eta)}{b(\eta)} \, \dot \eta \, \dot \sigma -   f'(\sigma) \, e^{2 \, f(\sigma)} \, \frac {a^2(\eta)} {b^2(\eta)}\, \,  [\, -\dot \eta^2 +  {\textstyle{\sum_ i}} (\dot x_i)^2 \, ] = 0,  \label{eq:sigmaeqn}
\end{equation}
here the dots represent the differentiation with respect to the affine 
parameter $\lambda$ (or an arbitrary parameter, for null geodesics) 
while primes denote differentiation of the functions 
with respect to their corresponding independent variables, $\eta$ or $\sigma$. The geodesic equations corresponding to $x$, $y$, and $z$ can be reproduced 
from equation (\ref{eq:xieqn}) and they all have an identical structure 
except for the constants $D_i$. The full geodesic equations in a spacetime 
corresponding to the metric (\ref{eq:cmetric}) are very difficult to solve 
analytically. However, the Eqs (\ref{eq:etaeqn})-(\ref{eq:sigmaeqn}) 
can be recast as the following dynamical system of first order, coupled differential equations,
\begin{eqnarray}
\dot \eta &=& \frac{e^{- \, f(\sigma)}}{a(\eta)}\sqrt{\epsilon + \frac{\chi^2}{b^2(\eta)}}, \label{eq:eta} \\
\dot x_i &=& \frac{D_i e^{-2f(\sigma)}}{a^2(\eta)},\label{eq:x} \\
\dot \sigma &=& \frac{\chi}{b^2(\eta)}, \label{eq:sigma} \\
\mbox{and} \hspace{.5cm} \dot\chi &=& - f'(\sigma) \left(\epsilon + \frac{\chi^2}{b^2(\eta)}\right).\label{eq:chi}
\end{eqnarray}
It will be interesting to see if one can analytically draw some useful 
conclusions about the behaviour of the geodesics by using the above set of 
equations. For the growing warp factor, $f'(\sigma) \sim \tanh (k\sigma)$ and 
if we expand this function about the location of the brane (i.e. $\sigma = 0$), the  dominant term in the neighbourhood of that point will be linear in 
$\sigma$. In such a scenario, for the timelike case ($\epsilon = 1$), the 
Eqs (\ref{eq:sigma}) and  (\ref{eq:chi}) resemble the following equations,
\begin{equation}
 \dot \sigma \sim \chi \,\times \, \mbox{positive terms}, \hspace{1cm} \mbox{and} \hspace{1cm} \dot\chi \sim \,-\, \sigma \times \mbox{positive terms}, \label{eq:argu} 
\end{equation}
which are qualitatively similar to a dynamical system representing simple 
harmonic 
motion (SHM) $\{\dot X = P, \dot P = - X\}$ (modulo a multiplicative factor). 
From 
this consideration, one can intuitively argue that the Eqs (\ref{eq:eta})-(\ref{eq:chi}) will have oscillatory solutions for $\sigma$ and $\chi$ (though, not {\em simple} harmonic). On the other hand, for a decaying warp factor, $f(\sigma) = -\log(\cosh \,k\sigma)$, the above set of equations given by (\ref{eq:argu}) lead to a system with 
exponential solutions. Thus, there should not be any oscillatory trajectories 
with decaying warp factors. 

From equation (\ref{eq:argu}), it is also evident that, for {\em any} 
growing warp factor one does not have oscillatory timelike 
trajectories. For example, in the Randall--Sundrum scenario, we have, $f(\sigma) \sim |\sigma|$ and $f'(\sigma)$ is therefore a step function, which cannot 
have an expansion with a term linear in $\sigma$. So there exist no 
oscillatory solution for this case. 

\begin{figure}[!ht]
\includegraphics[width = 5.5 in]{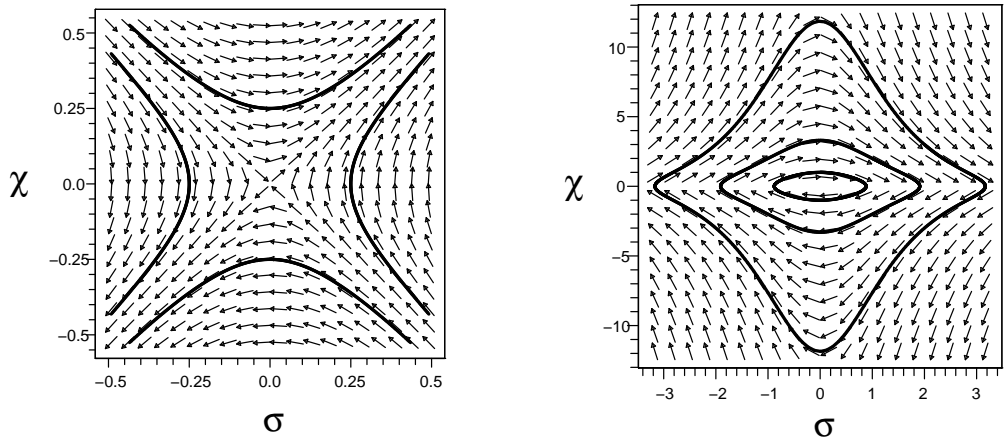}

\caption{The solid lines indicate different flows in the solution space of \{$\sigma$,$\chi$\} for different initial conditions, such as, in the left plot (for decaying warp factor): $\{\sigma(0),\chi(0)\} = (0.25,0), (0,-0.25), (-0.25,0), (0,0.25)$ (here $\lambda$ runs from -1.3 to 1.3),  and in the right plot (for growing warp factor): $\{\sigma(0),\chi(0)\} = (0,1), (1,2), (2,3)$ (here $\lambda$ runs from 0 to 6.5).}\label{fig:phaseplot1}
\end{figure}

In order to further analyse the behavior 
of trajectories  analytically, let us look at the specific case where $b(\eta)$ is a constant (say $b(\eta) = 1$). Since we are particularly interested about motion in the bulk, we choose, $\dot x_i = 0$ (as $x_i$ does not appear in the other three equations i.e. in  (\ref{eq:eta}), (\ref{eq:sigma}), and (\ref{eq:chi}). With these considerations, we arrive at following set of equations,
\begin{eqnarray}
\dot \eta &=& \frac{e^{- \, f(\sigma)}}{a(\eta)}\sqrt{\epsilon + \chi^2}, \label{eq:eta1}\\
\dot \sigma &=& \chi, \label{eq:sigma1} \\
\mbox{and} \hspace{.5cm} \dot\chi &=& -  f'(\sigma) \left(\epsilon + \chi^2\right) \label{eq:chi1}.
\end{eqnarray}
For timelike geodesics and growing warp factors, from Eqs (\ref{eq:sigma1}) and (\ref{eq:chi1}), we obtain,
\begin{equation}
\frac{d\sigma}{d\chi} = -  \frac{\chi}{(1 + \chi^2)\tanh\sigma} \hspace{.5cm}  \implies \hspace{.5cm} C_1\,{\rm sech}\,\sigma = \sqrt{1 + \chi^2}, \label{eq:sigma1chi1}
\end{equation}
where $C_1$ is an integration constant. Similarly, Eqs (\ref{eq:eta1}), (\ref{eq:chi1}) and (\ref{eq:sigma1chi1}) further lead to,
\begin{equation}
\frac{d\chi}{d\eta} = - 2\eta\, \sqrt{C_1^2 - 1 - \chi^2} \hspace{.5cm}  \implies \hspace{.5cm} \chi = \sqrt{C_1^2 - 1} \sin(C_2 - \eta^2),  \label{eq:chi1eta1} 
\end{equation}
where $C_2$ is another integration constant. Now combining the Eqs (\ref{eq:sigma1chi1}) and (\ref{eq:chi1eta1}), one easily obtains,
\begin{equation}
\sigma = {\rm sech}^{-1}\left[\frac{1}{C_1}\sqrt{1+ \big(C_1^2 - 1\big)\,\sin^2\big(C_2 - \eta^2\big)}\,\right].  \label{eq:sigma1eta1}
\end{equation}
The argument of the ``${\rm sech}^{-1}$'' function contains a sinusoidal part which guarantees an oscillatory behaviour of $\sigma(\eta)$, though its details depend on the parameters ($C_1$ and $C_2$) involved in the analysis.
It is worth mentioning here that even if we let $a(\eta)$ to be a constant, oscillations persist (only the factor $\sin^2(C_2 - \eta^2)$ gets replaced by $\sin^2(C_2 - 2\eta)$). This confirms that the oscillations are solely due to the presence of a growing warp factor. It has been checked that, for a decaying warp 
factor, no such analytic expression can be found.  
In fact, the Eqs (\ref{eq:eta}), (\ref{eq:sigma}) and (\ref{eq:chi}) form a dynamical system of their own, irrespective of equation (\ref{eq:x}). Unfortunately, with $\epsilon = 1$, this system does not have any fixed points for a finite value of $\eta$.
But, for a static extra dimension, we can investigate the solution space \cite{Strogatz} for the following set of two equations,
\begin{equation}   
\dot \sigma = \chi, \hspace{1cm}\mbox{and} \hspace{1cm} \dot \chi = \pm (1 + \chi^2)\tanh k\sigma , \label{eq:phasespace1} 
\end{equation}
where $+$ and $-$ signatures denote the decaying and growing warp factors respectively. It can easily be shown that the point $(\sigma,\chi) = (0,0)$ is a saddle point in the phase space for the decaying warp factor and a centre in case of a growing warp factor. Thus, we can expect oscillatory solutions for $\sigma(\lambda)$ and $\chi(\lambda)$ in the latter case (i.e. for a growing warp factor). We present the phase space plots of $\sigma$ and $\chi$ with two different warp factors  in the Fig. \ref{fig:phaseplot1}.
A decaying warp factor only helps a particle to move away from the brane, except for the initial conditions satisfying $\sigma + \chi = 0$ (where the particle tends toward the location of the brane with an ever-decreasing velocity). This essentially means that only for the abovementioned initial conditions, particles will end up accumulating near the brane with increasing affine parameter. Particles with initial condition $\sigma - \chi = 0$ will be repelled away from the brane. There are two other kinds of trajectories.  The trajectories in the quadrants ($\sigma + \chi > 0, \sigma - \chi < 0$) and ($\sigma + \chi < 0, \sigma - \chi > 0$) may cross the brane at some point of time (depending on the initial conditions), and the trajectories in the quadrants ($\sigma + \chi > 0, \sigma - \chi > 0$) and ($\sigma + \chi < 0, \sigma - \chi < 0$) will bypass the location of the brane (though they may come very close, depending on the initial conditions). On the other hand, the closed curves, in the case with a growing warp factor, indicate that, $\sigma(\lambda)$ and $\chi(\lambda)$ have an oscillatory behaviour. One can also notice that the nature of the oscillations depend heavily on the initial conditions. A particle, initially on or close to the brane and having a small velocity, executes an almost sinusoidal oscillation. This oscillatory pattern will tend towards a square wave profile as the distance of the launching point from the brane, or the initial velocity, increases. The phase portraits in Fig. \ref{fig:phaseplot1} are for a static extra dimension. We expect to see a similar solution space for a dynamic extra dimension ($b(\eta)$) too when $\eta \rightarrow \infty$ because according to our choices $b(\eta)$ gets stablised to a constant value for large $\eta$.  The closed trajectories for a static extra dimension and a growing warp factor would become limit cycles in phase space when we have a dynamic extra dimension.

One may also note that for null geodesics, the eigenvalues of the linearised system at the origin in the phase space vanishes. It is therefore not possible to comment on the solution space for this case. 

\subsection{Numerical evaluations}
\begin{figure}[!ht]
\includegraphics[width = 2.4 in]{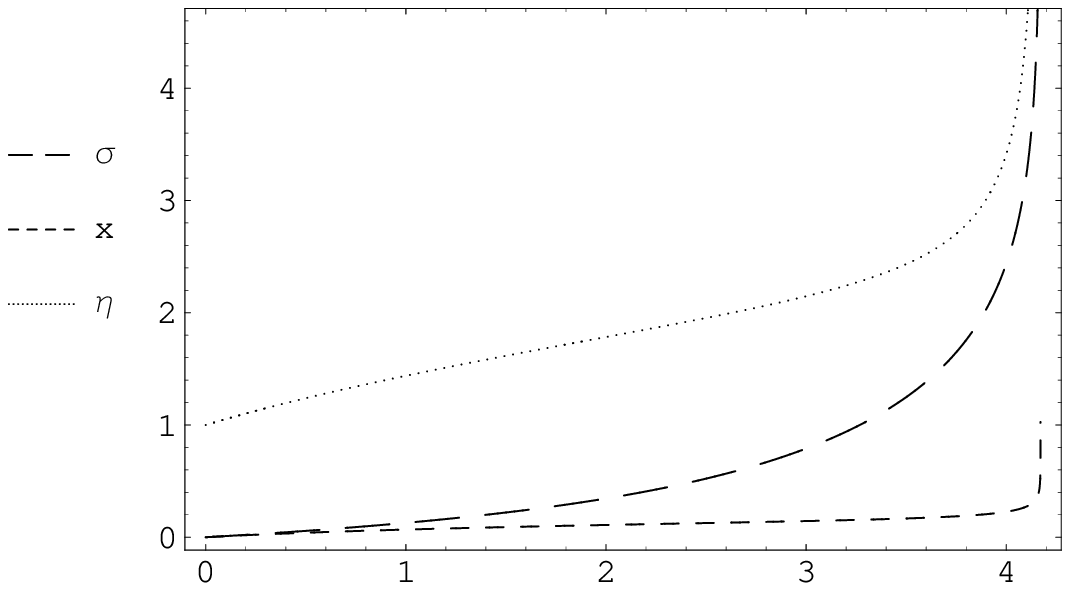}\hspace{.5 cm}
\includegraphics[width = 2.4 in]{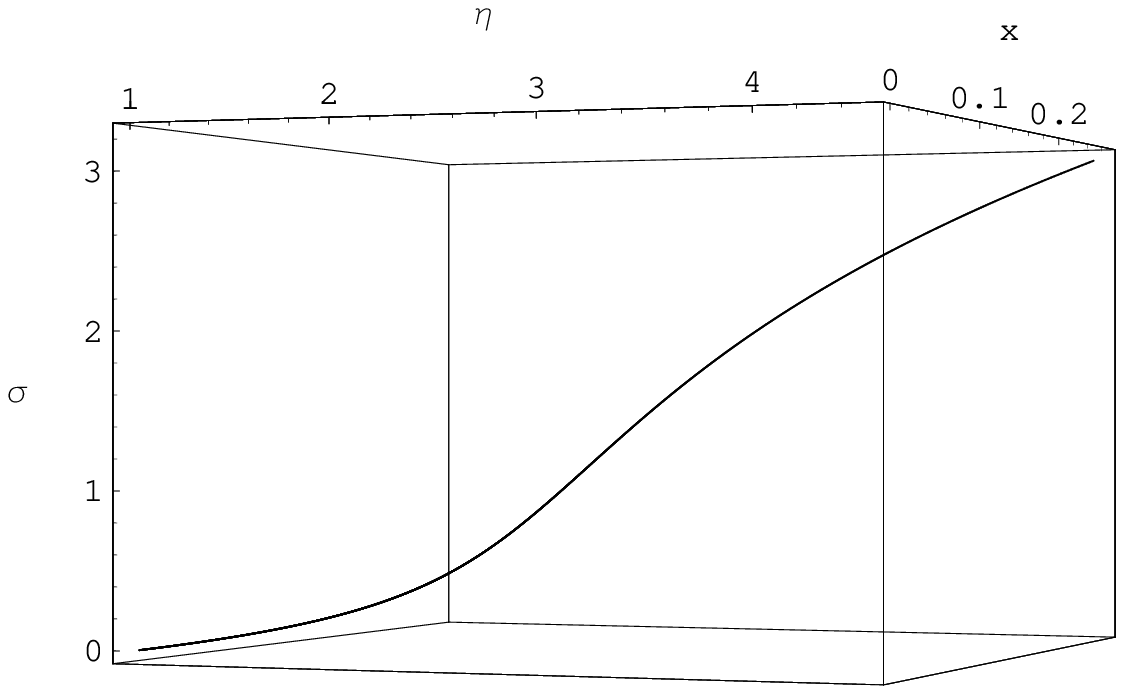}
\fbox{\includegraphics[width = 2 in]{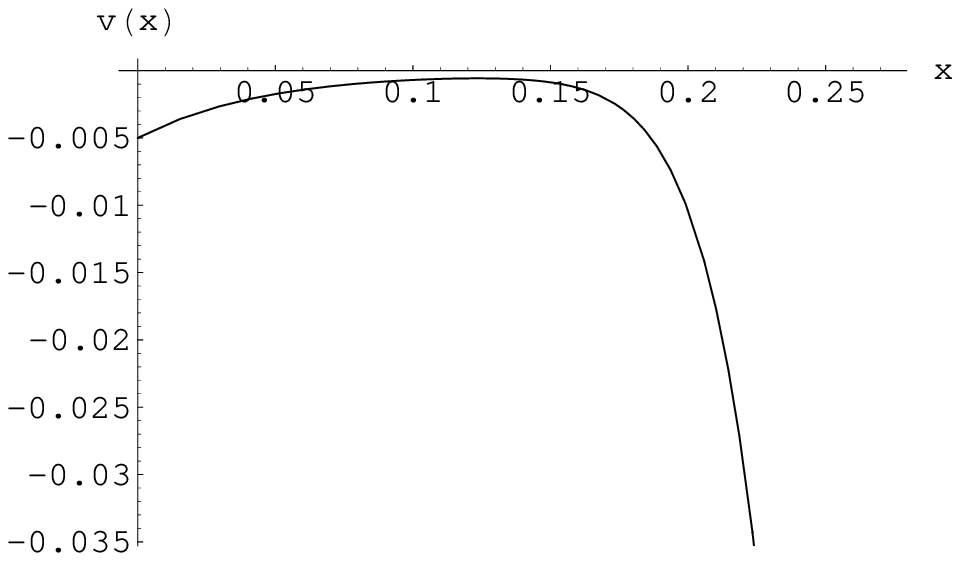}}\hspace{1 cm}
\fbox{\includegraphics[width = 2 in]{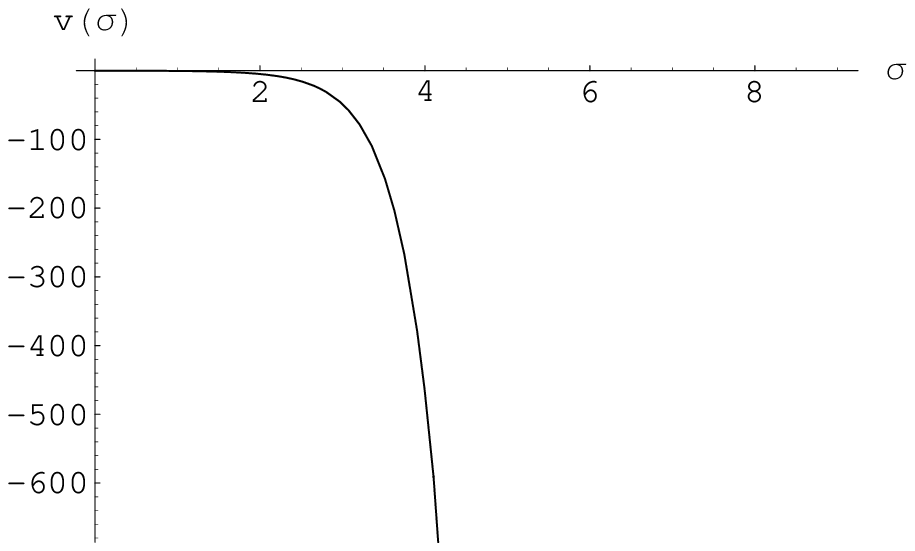}}
\caption{Timelike geodesics for set (A) with a decaying warp factor.} \label{NumGeo3_1}
\end{figure}

\begin{figure}[!ht]
\includegraphics[width = 2.4 in]{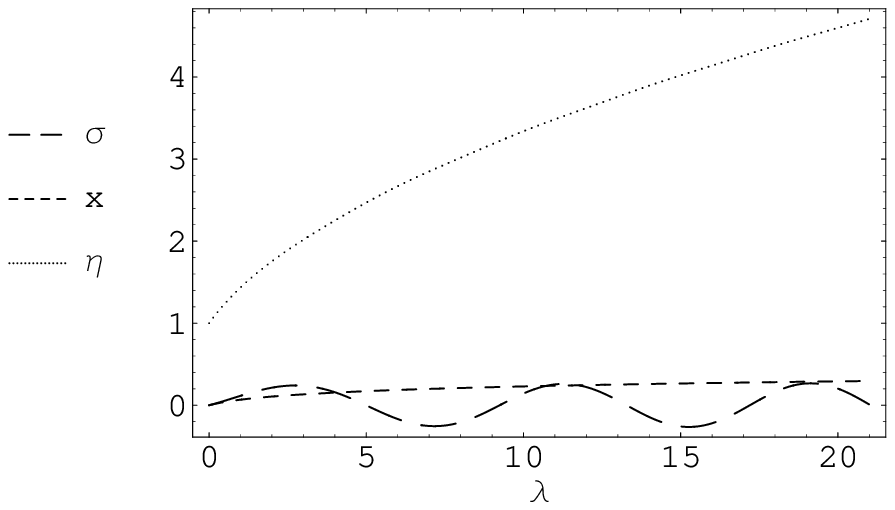}\hspace{.5 cm}
\includegraphics[width = 2.4 in]{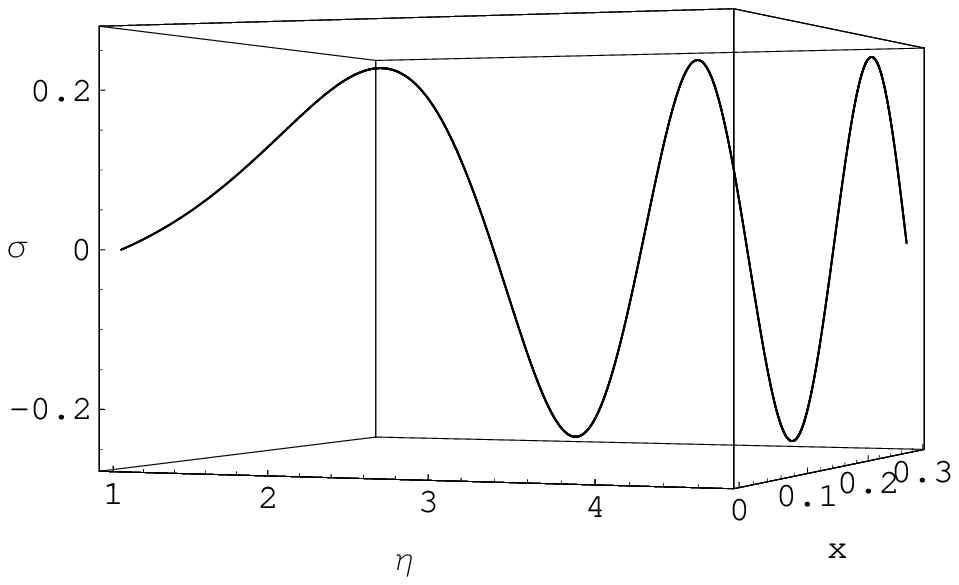}
\fbox{\includegraphics[width = 2 in]{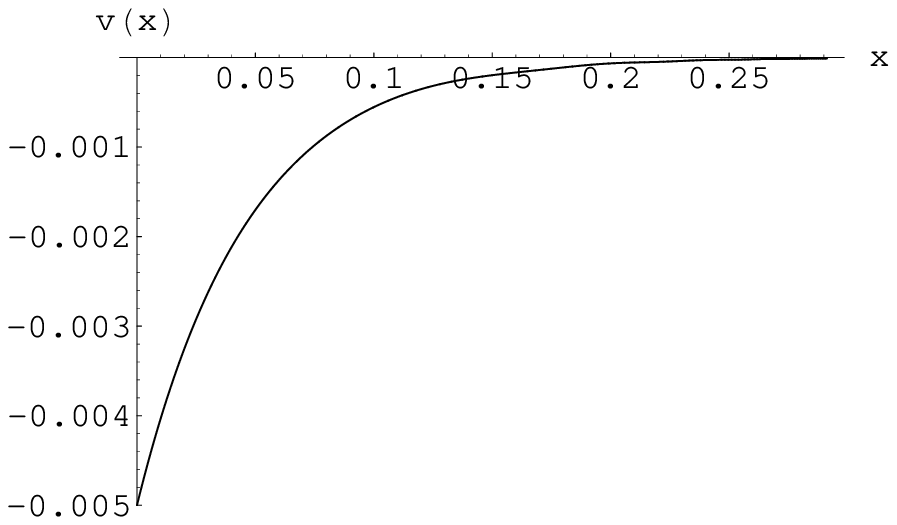}}\hspace{1 cm}
\fbox{\includegraphics[width = 2 in]{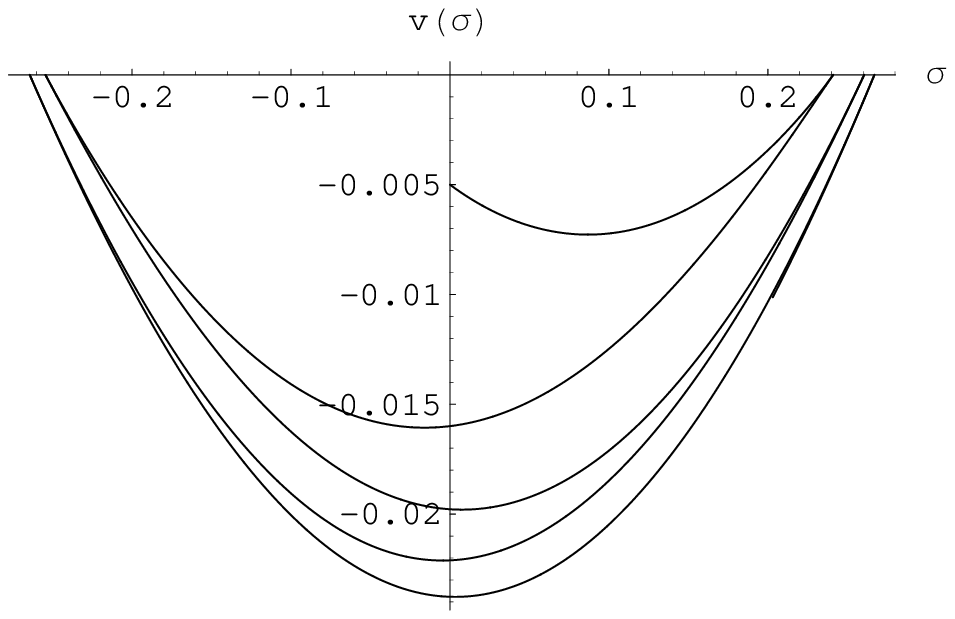}}
\caption{Timelike geodesics for set (A) with a growing warp factor.} \label{NumGeo4_1}
\end{figure}

\begin{figure}[!ht]
\includegraphics[width = 2.4 in]{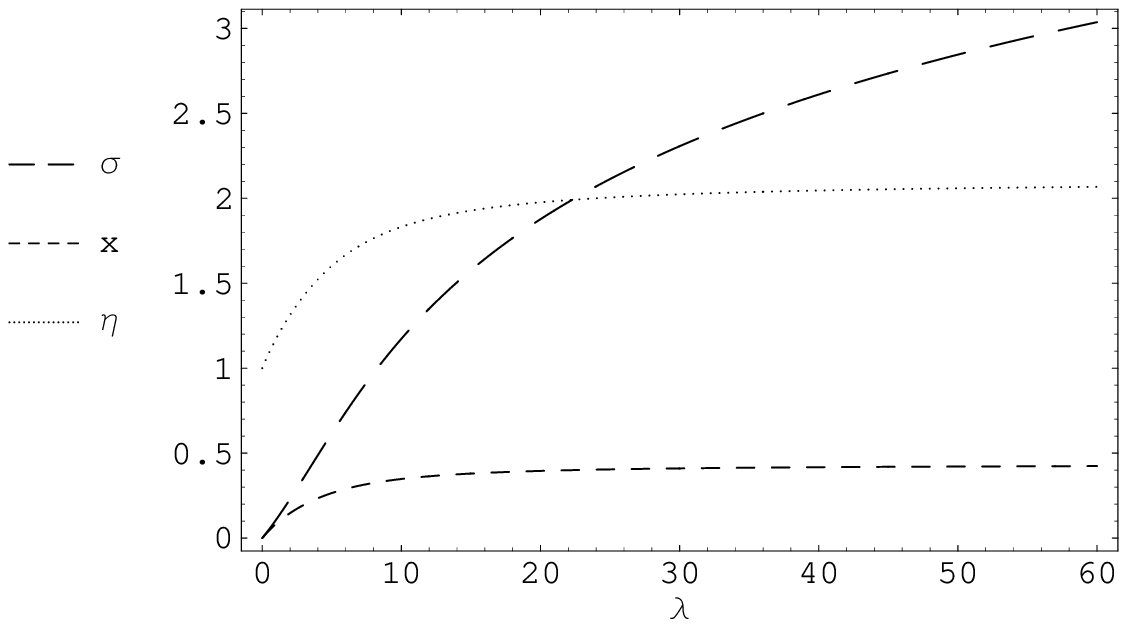}\hspace{.5 cm}
\includegraphics[width = 2.4 in]{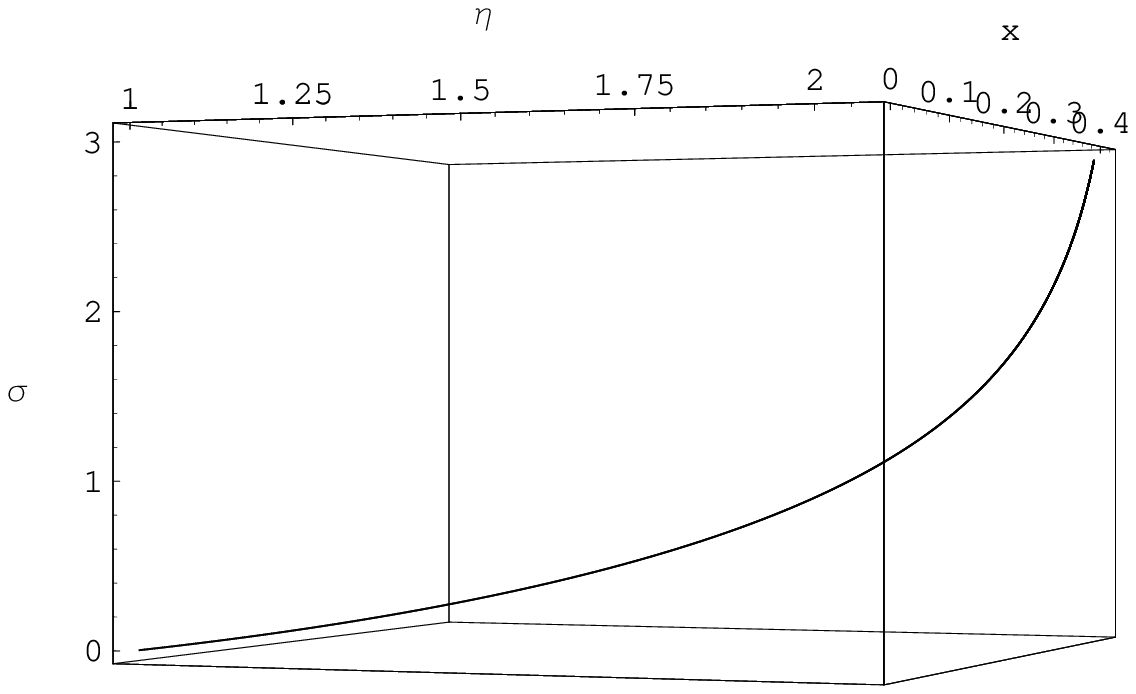}
\fbox{\includegraphics[width = 2 in]{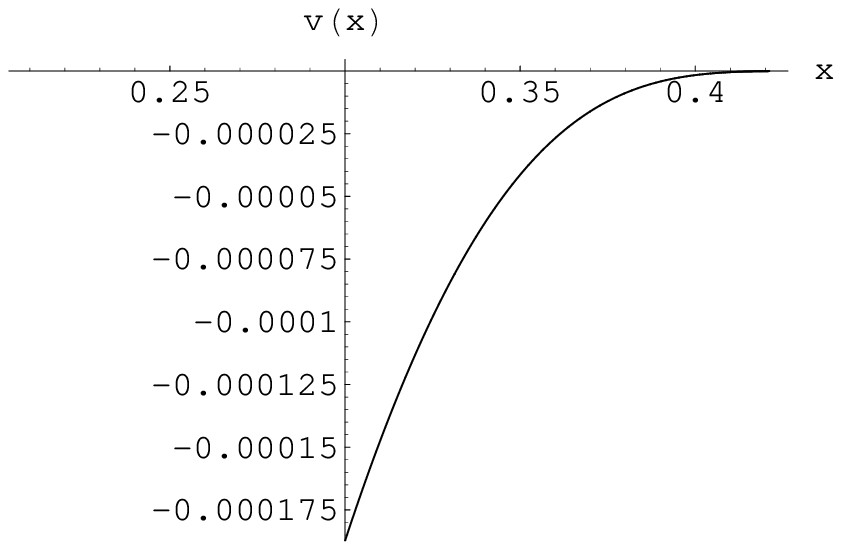}}\hspace{1 cm}
\fbox{\includegraphics[width = 2 in]{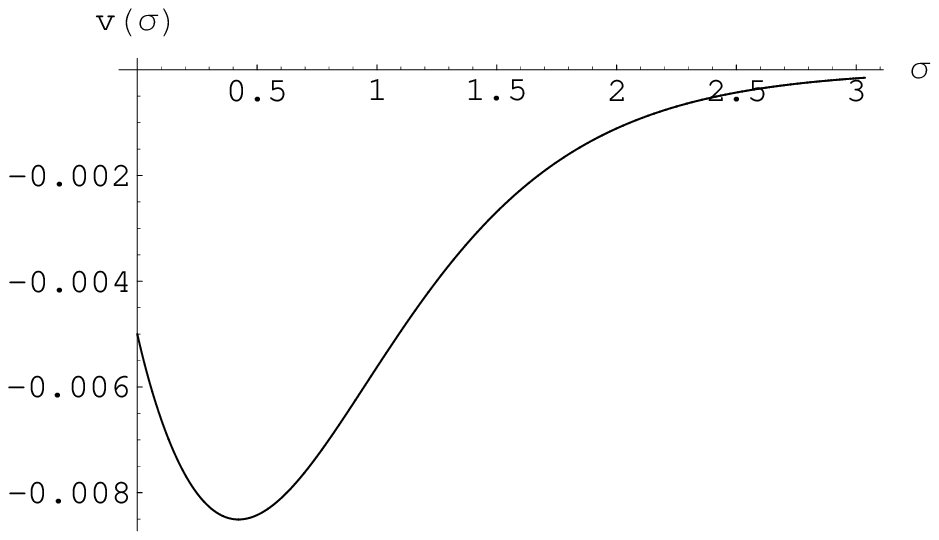}}
\caption{Null geodesics for set (A) with a growing warp factor.} \label{NumGeo4_2}
\end{figure}

\begin{figure}[!ht]
\includegraphics[width = 2.4 in]{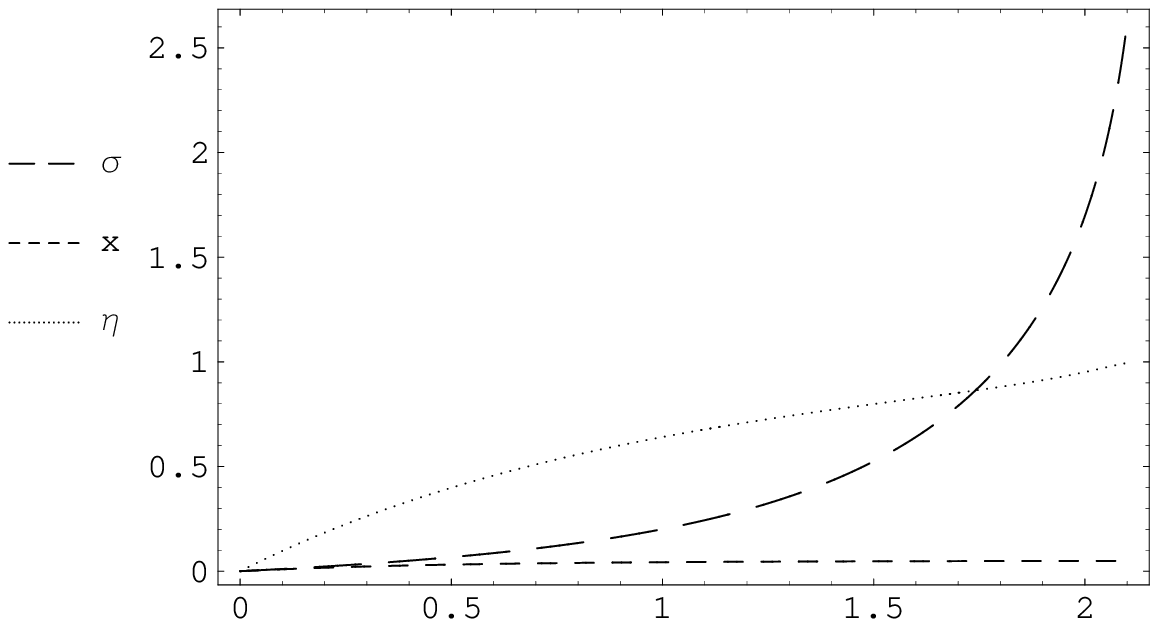}\hspace{.5 cm}
\includegraphics[width = 2.4 in]{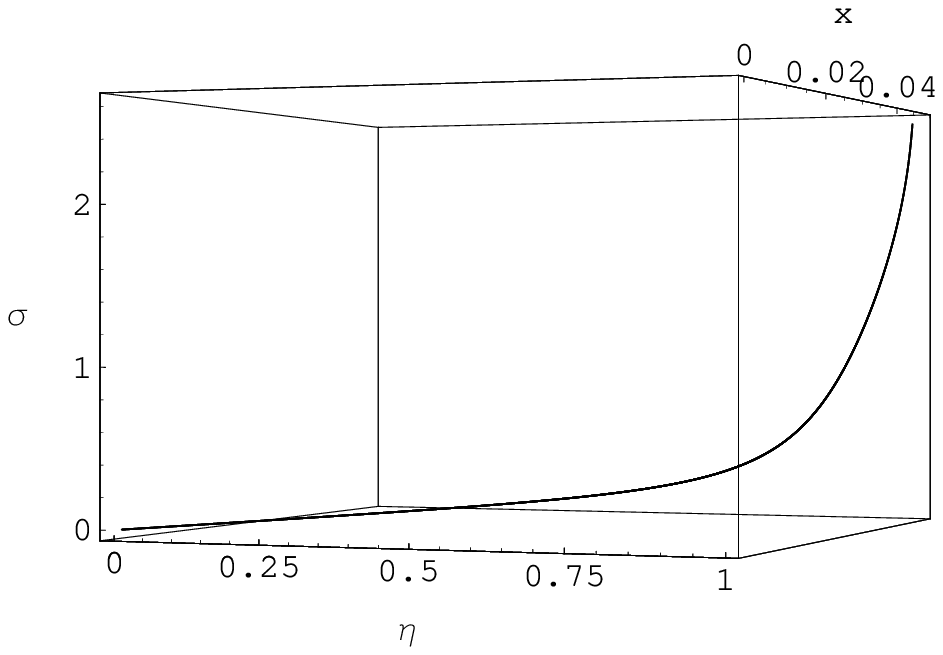}
\fbox{\includegraphics[width = 2 in]{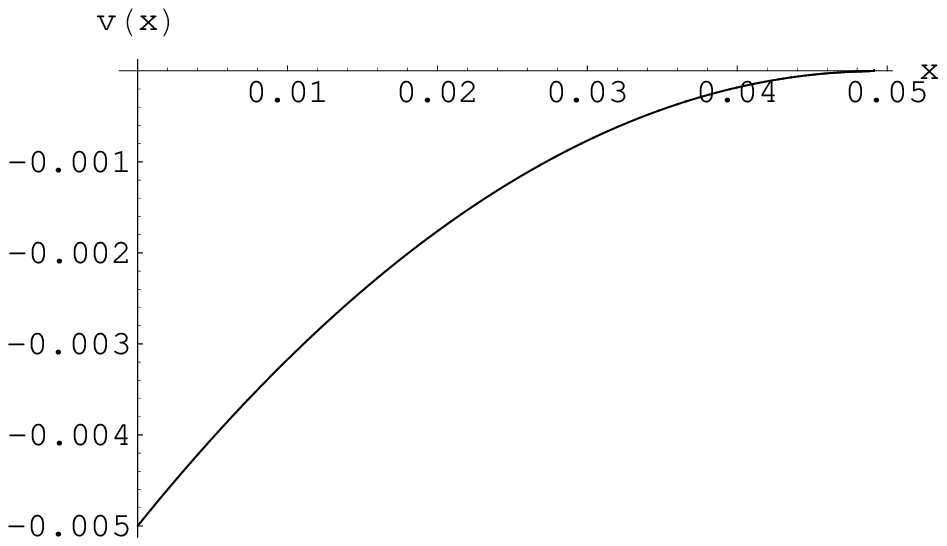}}\hspace{1 cm}
\fbox{\includegraphics[width = 2 in]{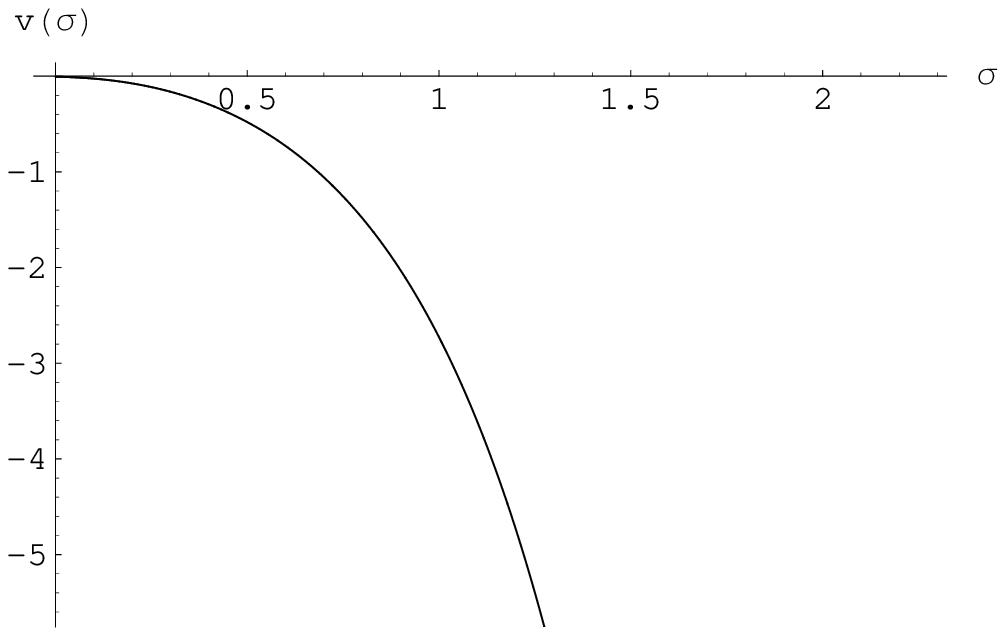}}
\caption{Timelike geodesics for set (B) with a decaying warp factor.} \label{NumGeo7_1}
\end{figure}

\begin{figure}[!ht]
\includegraphics[width = 2.4 in]{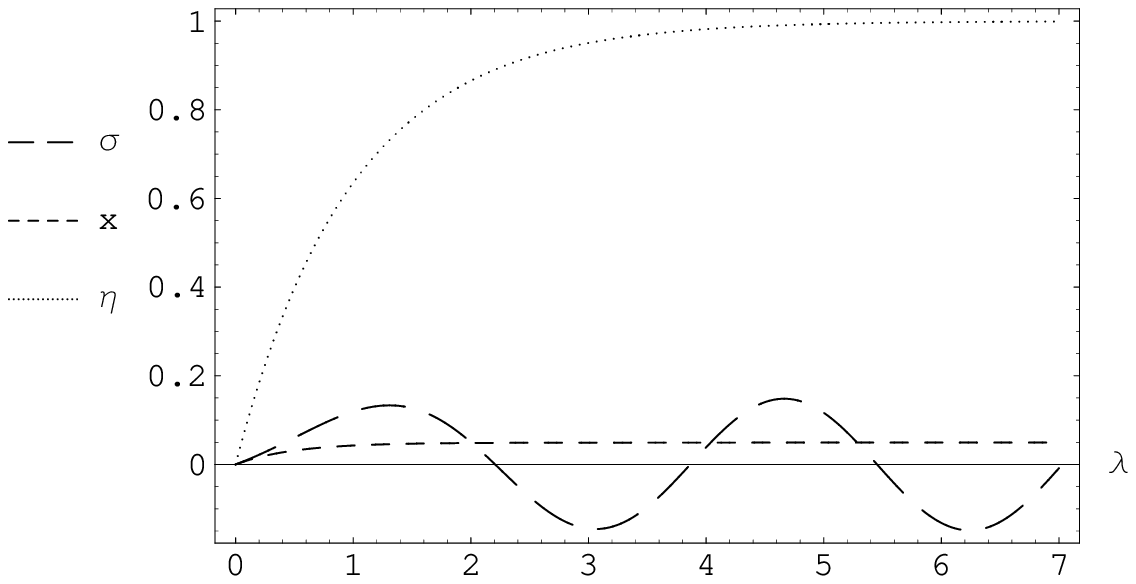}\hspace{.5 cm}
\includegraphics[width = 2.4 in]{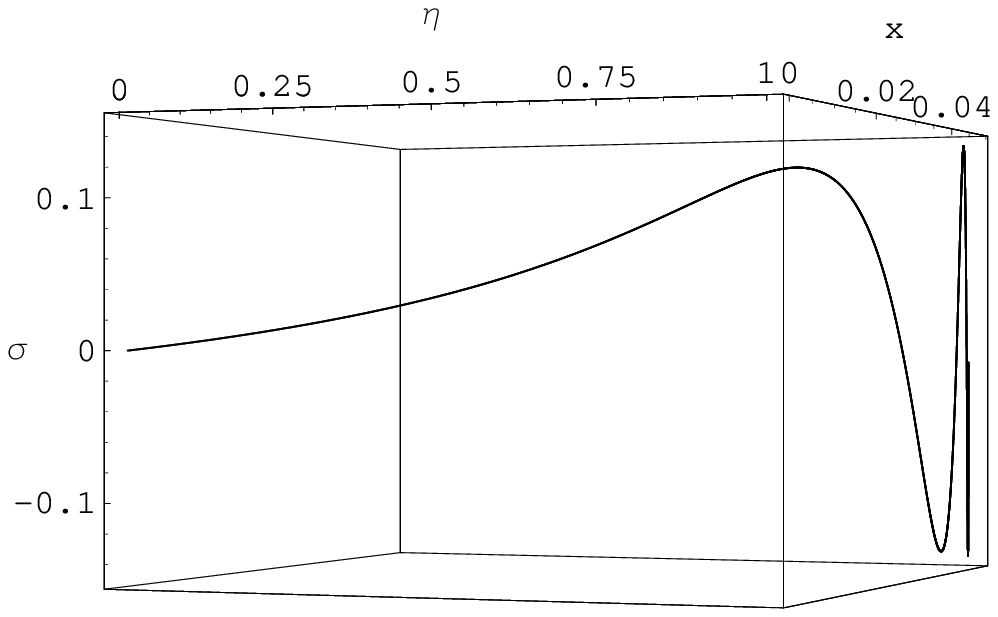}
\fbox{\includegraphics[width = 2 in]{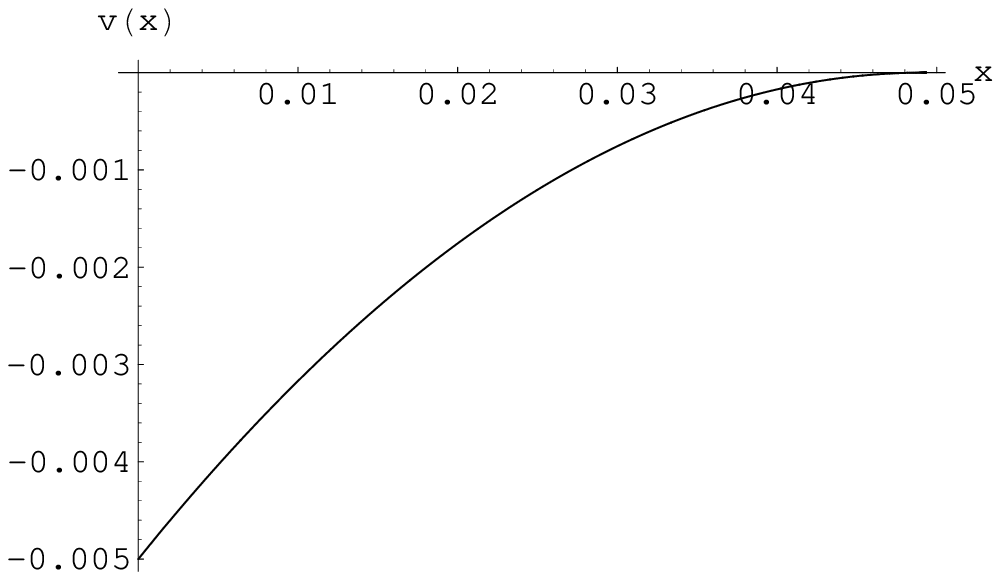}}\hspace{1 cm}
\fbox{\includegraphics[width = 2 in]{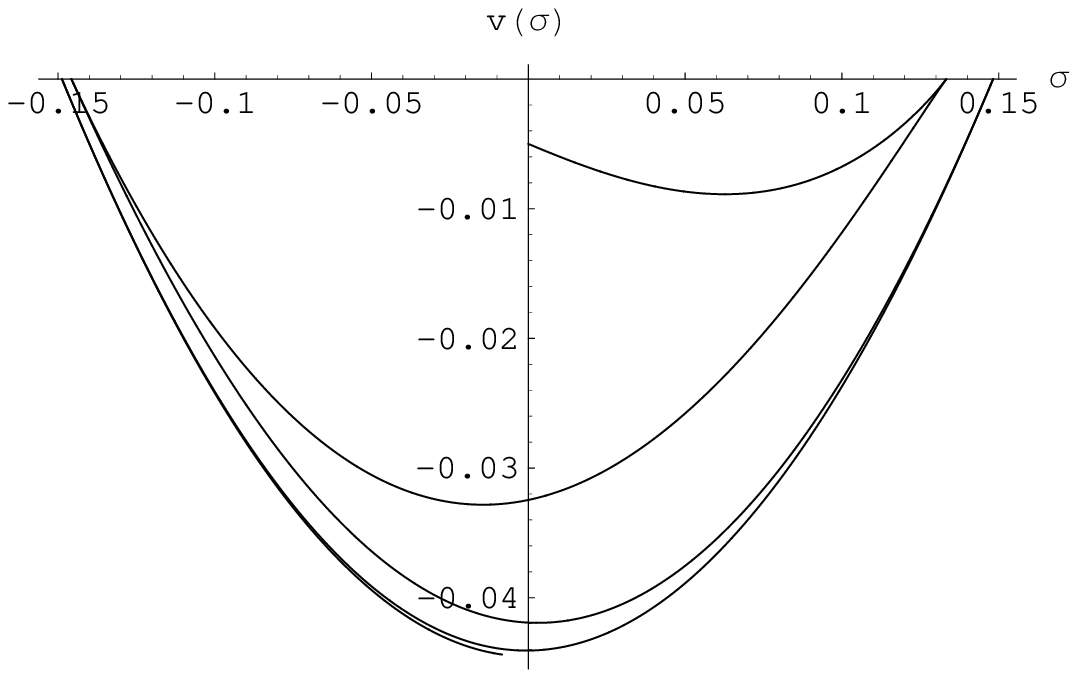}}
\caption{Timelike geodesics for set (B) with a growing warp factor.} \label{NumGeo8_1}
\end{figure}

\begin{figure}[!ht]
\includegraphics[width = 2.4 in]{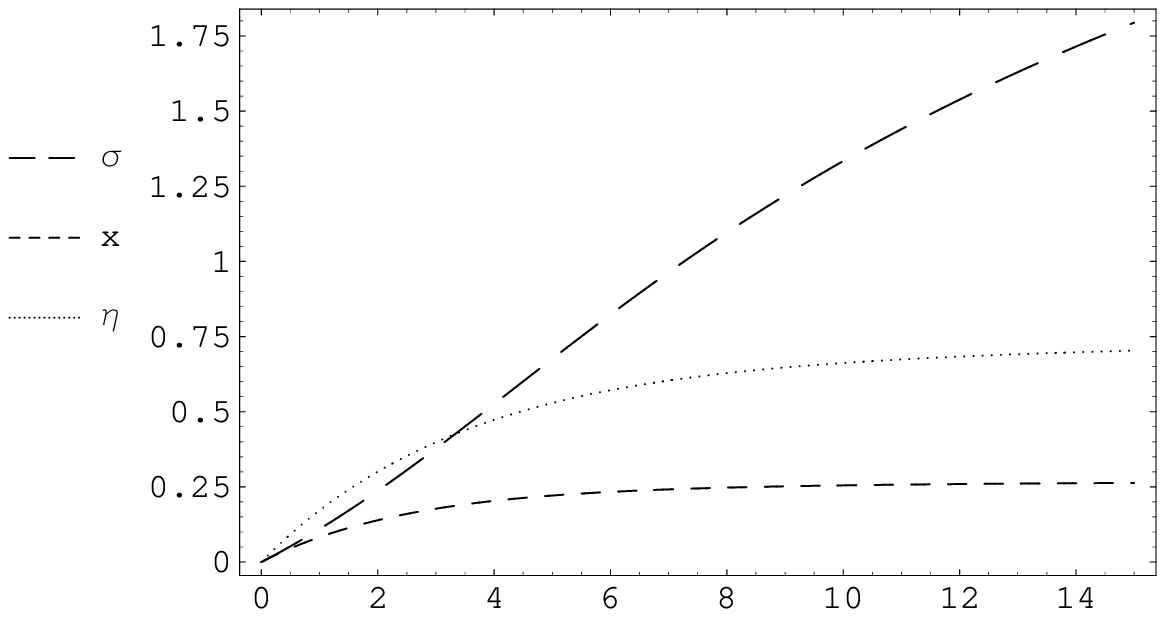}\hspace{.5cm}
\includegraphics[width = 2.4 in]{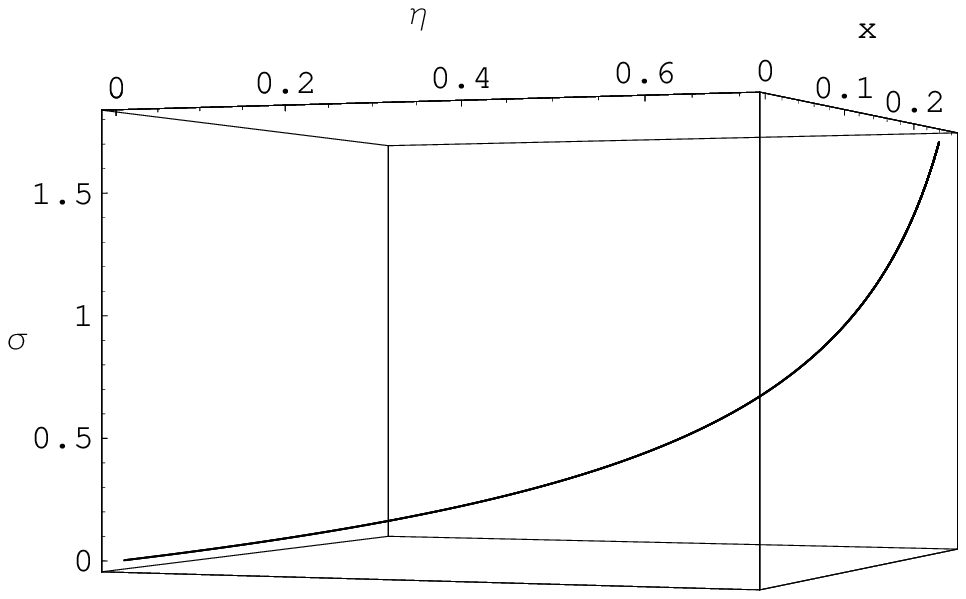}
\fbox{\includegraphics[width = 2 in]{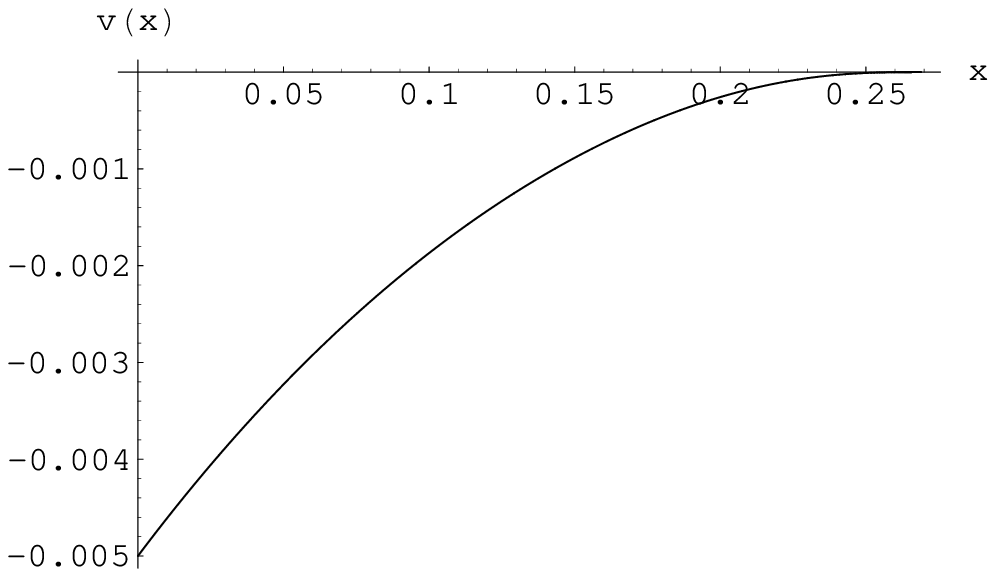}}\hspace{1 cm}
\fbox{\includegraphics[width = 2 in]{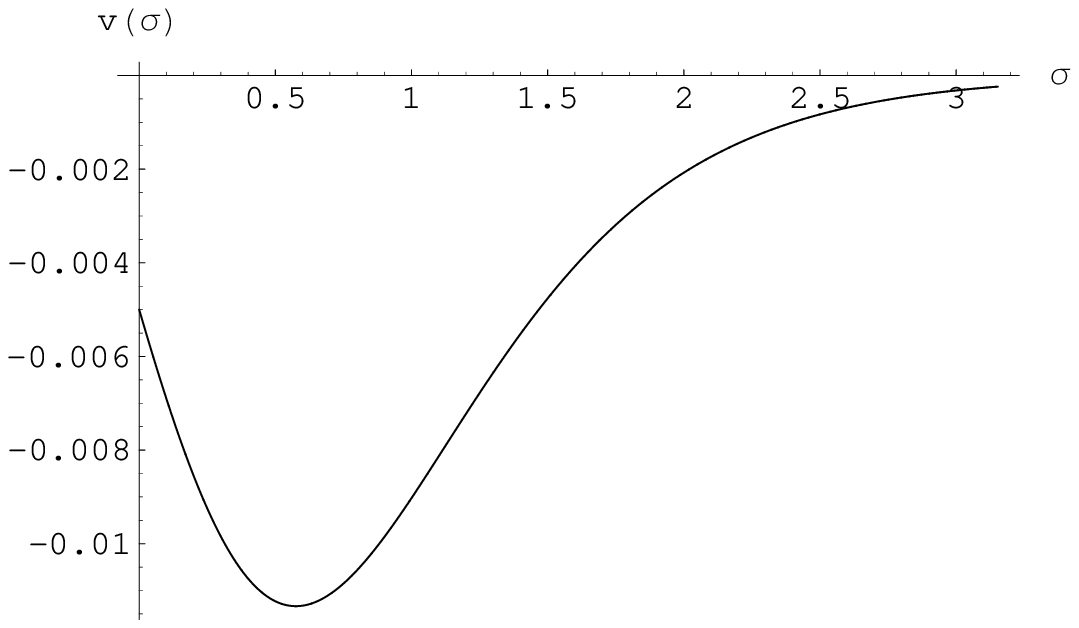}}
\caption{Null geodesics for set (B) with a growing warp factor.} \label{NumGeo8_2}
\end{figure}

 The full geodesic Eqs (\ref{eq:etaeqn})-(\ref{eq:sigmaeqn}) for the 
null and timelike geodesics can obviously be solved numerically. We present 
some features of these geodesics, graphically, with the set (A) and (B) for the 
scale factors and with the growing and decaying warp factors of the 
thick brane model. All of these cases  are presented in Figs. \ref{NumGeo3_1} -\ref{NumGeo8_2} where the figure captions are self--explanatory and 
describe the corresponding results.  
We have shown the solutions for timelike and null geodesics in the case 
with a growing warp factor. Here, we have observed remarkable differences 
between the null and timelike cases (see Figs. \ref{NumGeo4_1}, \ref{NumGeo4_2}, \ref{NumGeo8_1} and \ref{NumGeo8_2}). Otherwise, where the differences 
between the results for timelike and null cases are small (i.e. results 
remain qualitatively similar for the case of a decaying warp factor), we 
present only the timelike geodesics (see Figs. \ref{NumGeo3_1} and 
\ref{NumGeo7_1}). The initial conditions for our numerical evaluations are 
chosen as $\dot x = \dot y = \dot z = \dot \sigma = 0.1$ at $\lambda =0$ 
and $\dot \eta $ at $\lambda =0$ is calculated by using the null and timelike 
constraints which differs between models.

We have observed that for a growing warp factor, the timelike geodesics have
 an oscillatory behaviour for both the cases (i.e. set (A) and (B)) as 
presented in Figs. \ref{NumGeo4_1} and \ref{NumGeo8_1} respectively. Such an 
oscillatory behaviour is absent in the trajectories of null geodesics in all 
the cases. This essentially means that massless particles are free to escape 
into the bulk. The oscillatory behaviour of the timelike geodesics for a 
growing warp factor also manifests itself in terms of the potential 
$V(\sigma)$ as shown in Figs. \ref{NumGeo4_1} and \ref{NumGeo8_1} for set (A) 
and set (B) respectively. The fact which is worth mentioning about these 
trajectories is that a massive particle never moves away to an infinite 
distance from the location of the brane. It seems that the trajectories are 
automatically localised on or near the brane in the presence of a growing warp 
factor. The potential in this case is in fact that of an anharmonic oscillator 
where both the amplitude and frequency of the oscillations are monotonically 
increasing and converging towards a limiting value with an increasing 
$\lambda$. We will again return to the interpretation of this feature in terms 
of the warp factor and the scale of the extra dimension, in the Section 
\ref{comparisonavp}. Our numerical findings in this section tally well 
with the results found earlier in Section III, via the analysis of an autonomous
dynamical system derived from geodesic equations.

\section{Effects of warping and time dependence in extra dimension scale 
and confinement of test particles} \label{comparisonavp}

In order to figure out the effects of warping and a time dependent extra dimension, we will compare the generalised braneworld scenario (discussed in the previous section) with the situations where we do not have any warp factor or a time dependence in the extra dimension scale. Let us first consider, the metric (\ref{eq:cmetric}) without the warping factor. This leads to the following constraint,
\begin{equation}
a^2(\eta)\,  [- d\eta^2 + d{\bf X}^2 ]\, + b(\eta)^2\dot \sigma^2 + \epsilon = 0. \label{eq:conmetric_comp1}
\end{equation}
Now $\sigma$ becomes cyclic in the geodesic Lagrangian and gives us $V_x(x) = - D^2/(2 a^4)$ and $V_{\sigma}(\sigma) = - P^2/(2b^4)$ which eventually lead to,
\begin{equation}
\lambda = \int{\frac{a d\eta}{\sqrt{\left(\epsilon + \frac{D^2}{a^2} + \frac{P^2}{b^2}\right)}}}. \label{eq:eqn_comp1}
\end{equation}
With a nonzero $P$, we could not find an analytic solution for Eq. \ref{eq:eqn_comp1} (though it is exactly solvable for $P = 0$). 
We have therefore solved the geodesic equations for $P \neq 0$ numerically. 

In the case where $b(\eta)$ is constant, the constraint (\ref{eq:conmetric}) gives,
\begin{equation}
\frac{d\eta}{d\lambda} = \frac{e^{-f}}{a} \, \sqrt{\left(\epsilon + \dot\sigma^2 + \frac{D^2}{a^2e^{2f}} \right)}. \label{eq:eqn_comp2}
\end{equation}
The Eq. (\ref{eq:eqn_comp2}) is also not analytically solvable except for the 
case with $D = 0$. As above, we solve the geodesic equations numerically 
for the general situation when $D \neq 0$. 

\begin{figure}[!ht]
\fbox{\includegraphics[width = 2 in]{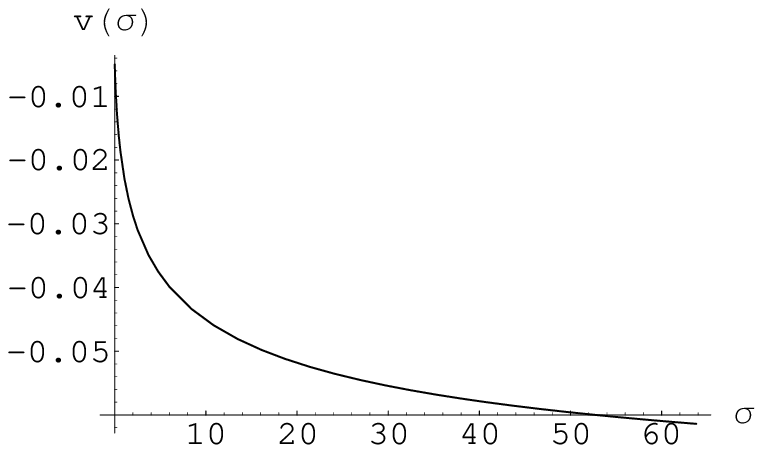}}\hspace{.5 cm}
\includegraphics[width = 2.4 in]{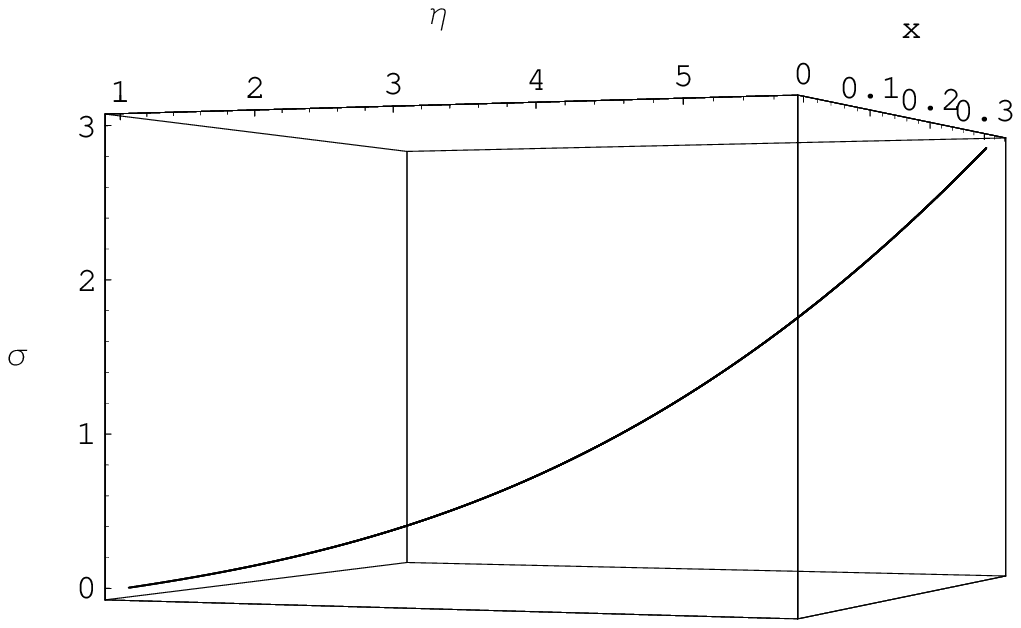}

\fbox{\includegraphics[width = 2 in]{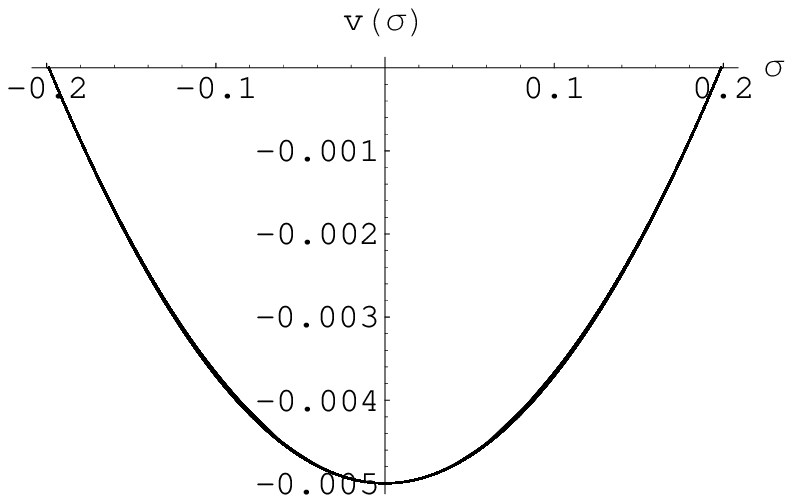}}\hspace{.5 cm}
\includegraphics[width = 2.4 in]{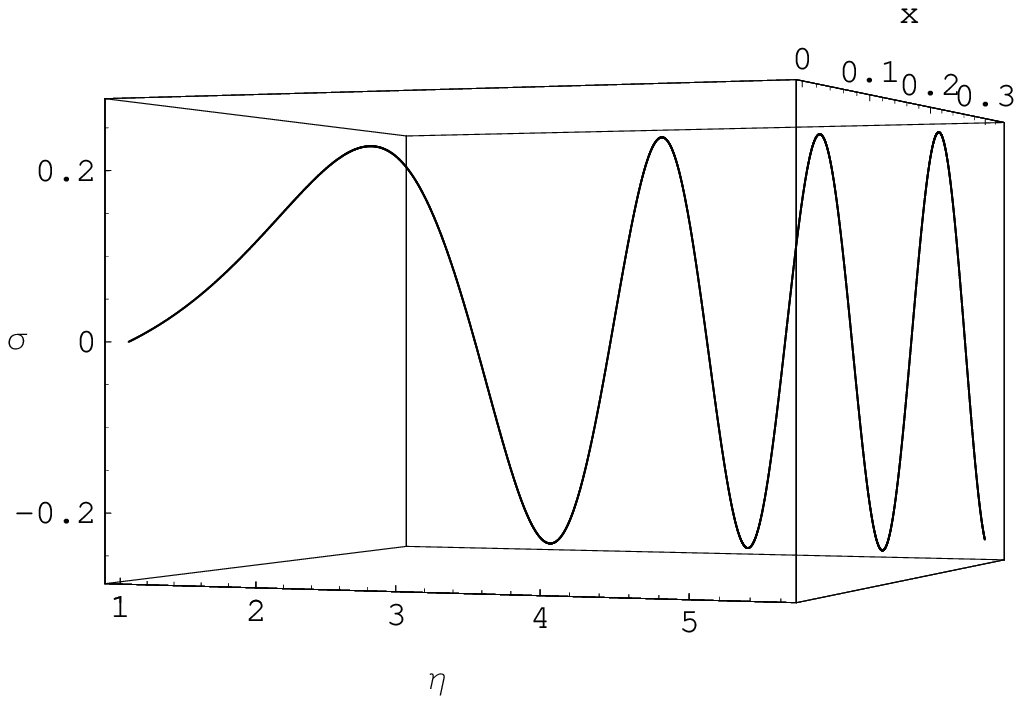}
\caption{Plots in the upper and lower rows show the timelike geodesics for set (A) (with $b(\eta) = constant$) for models without any warping factor and with growing warp factor respectively.} \label{fig:comparing}
\end{figure}

In order to see the specific 
differences between the general case (eg.  timelike 
geodesics for set (A) with growing warp factor as shown in Fig. \ref{NumGeo4_1}), and two of its subcases (one without a warping factor and the other with 
a static extra dimension), the plots of the potential $V_{\sigma}(\sigma)$ 
and the trajectories for both of these two subcases are presented in 
Fig. \ref{fig:comparing}. Here, we have considered $D=1$ and $P=1$ for the 
purpose of numerical computations. 
The differences between the two subcases are quite
clearly visible. We have also worked through other similar cases, which we
do not mention here.
It is clear that the oscillatory behaviour is due to the presence 
of a growing warp factor (as already argued in Section III). If we compare Fig. \ref{fig:comparing} with Fig. \ref{NumGeo4_1}, where scale of the extra dimension is dynamic, the increase in the amplitude and frequency of the oscillations of $\sigma$, may be understood as an effect of the decay in the scale of the 
extra dimension $b(\eta)$. In fact, the scale $b(\eta)$, as mentioned before, 
can be viewed as the length of an anharmonic oscillator which tends to 
become a harmonic one with an increasing value of the affine parameter. 
In other words, this problem is equivalent to a problem of a particle in 
a one dimensional box whose size is decreasing and tending to a finite value. 
This is exactly how one can understand the increase in the amplitude of 
the oscillation in $\sigma(\lambda)$, in Fig. \ref{NumGeo4_1}, and its 
eventual stabilisation to a limiting value. Such features are also evident from the behaviour of the corresponding  potentials as given in the Figs. \ref{NumGeo4_1} and \ref{fig:comparing}. 

Let us try to understand the above-mentioned confinement from a different point of view.
It was shown in \cite{seahra}, that in the case of a five dimensional bulk 
with a static extra dimension and a growing warp factor (which is indeed a solution of five dimensional Einstein equations in presence of a negative bulk 
cosmological constant), we have confinement of test particles. 
Further, in \cite{seahra}, it was shown 
that confinement will be achieved if the following energy condition holds,
\begin{equation}
({}^0 {\hat R}_{AB}e^A_{\alpha}e^B_{\beta} - {}^0 R_{\alpha \beta})u^{\alpha}u^{\beta} > 0, \hspace{1cm}A,B = 0,1,2,3,4; \hspace{0.5cm}\alpha,\beta = 0,1,2,3
\end{equation}  
where, $e^A_\alpha = \frac{\partial x^A}{\partial y^\alpha}$ are the four basis vectors and $\{x^A\}$ and $\{y^\alpha\}$ are five and four dimensional coordinate systems respectively. $\hat R_{AB}$ and $R_{\alpha \beta}$ are Ricci tensors in the bulk and on the $\sigma = const.$ hypersurface respectively.
The above condition implies that the local gravitational density of 
five dimensional bulk matter, 
as measured by an observer freely falling along a hypersurface $\sigma = 0$ 
(or the brane), has to be greater than the gravitational density of the 
effective four dimensional matter. For the five dimensional geometry given by Eq. \ref{eq:cmetric}, the above constraint leads to
\begin{equation}
\frac{f'' + 4f'^2}{a^2b^2} + \left(\frac{\dot a \dot b}{a b} - \frac{\ddot b}{b}\right){u^0}^2 + \frac{\dot a \dot b}{a b}\displaystyle\sum_{i=1}^3 {u^i}^2 > 0. \label{eq:conf_cond1}
\end{equation}  
It may be noted that for a static extra dimension, i.e. when $b(\eta) \sim constant$, the following condition,
\begin{equation}
f'' + 4f'^2 > 0  \label{eq:conf_cond2}
\end{equation}   
is sufficient for confinement of test particles. This condition is clearly satisfied for the functional form of the growing warp factor we have considered.

 Now the constraint Eq. \ref{eq:conf_cond1} is always satisfied for a monotonically growing warp factor with a growing $a(\eta)$ and a growing but decelerating $b(\eta)$. In that case, if $b(\eta)$ does not stabilise, then as $\eta \rightarrow \infty$, the depth of the potential well tends to zero (as $b(\eta)$ appears in the denominator of Eq. \ref{eq:vs}). As a result, confined trajectories will become more and more unstable against perturbations and particles should disappear from the brane. This observation should rule out the models with growing and divergent $b(\eta)$, for example, some of the power law solutions (for $b(t)$) found in \cite{sols}, in presence of different bulk matter fields as such. On the other hand, if $b(\eta)$ grows but stabilises to a finite value as $\eta \rightarrow \infty$, trajectories are confined, though become less strongly bound as the depth of the potential decreases (towards a limiting value). 

The last two terms in the left hand side of \ref{eq:conf_cond1} are always negative for a growing $a(\eta)$ and a decaying $b(\eta)$. 
Therefore, it is obvious that the inequality \ref{eq:conf_cond1} will not 
always hold for any set of metric functions (though it happens to be the case for the metric functions we choose). However, one can argue that as $b(\eta)$ 
stabilises with $\eta \rightarrow \infty$, constraint \ref{eq:conf_cond1} converges toward constraint \ref{eq:conf_cond2}. It is thus implied 
that confinement will eventually be achieved during the course of cosmological evolution. The trajectories, in this case, will also be more strongly bound as the depth of the potential well increases. It is worth mentioning here that, if $b(\eta)$ decays to a zero value, the depth the potential \ref{eq:vs} tends to infinity resulting in an ever-growing amplitude of oscillation which eventually diverges. This aspect may help us rule out models with the $b(\eta)$ decaying to zero (instead of stabilising to a finite limiting value).

One may also find a constraint on the warp factor, for confinement, in a static bulk in the following simple way. For the test particles (moving along the extra dimension) to be confined somewhere in the bulk, their geodesic potential should have a minimum at that point. Now the corresponding geodesic potential is given by 
\begin{equation}
V(\sigma) = -\frac{C e^{-2f(\sigma)} - \epsilon}{2},  \label{eq:geo_pot}
\end{equation} 
where $C$ is, a positive constant, dependent on the initial conditions. $V(\sigma)$ should have a minima at $\sigma = \sigma_0$, where $f'(\sigma_0)=0\ \mbox{and}\ e^{-2f(\sigma_0)}\ne 0$, if the following condition holds
\begin{equation}
f''(\sigma_0) > 0  \label{eq:conf_cond},
\end{equation}
which means the $f(\sigma)$ must be a monotonically growing function of 
$\sigma$ about the location where the potential has a minimum. For $f(\sigma) = \log(\cosh \sigma)$, $V(\sigma) = -\frac{C sech^2\sigma - \epsilon}{2}$. The minimum value of this potential for timelike geodesics is $-\frac{C - 1}{2}$ (where $C \ge 1$) at $\sigma = 0$ and it reaches its maximum value $+ \frac{1}{2}$ as $\sigma \rightarrow \pm\infty$. Therefore the zero energy trajectories are bounded within the $\sigma$ values which can be obtained by equating $V(\sigma) = 0$. This leads to 
\begin{equation}
f(\sigma) = \log \sqrt{C}.   \label{eq:bound}
\end{equation}
We can therefore explain the confinement of massive particles in the context of Figs \ref{NumGeo4_1} and \ref{NumGeo7_1}. On the other hand, for null geodesics, the zero energy  trajectories always have runaway features (Figs. \ref{NumGeo4_2} and \ref{NumGeo8_2}). 

We note that whenever Eq. \ref{eq:conf_cond} is true, the Eq. \ref{eq:conf_cond2} is automatically satisfied. This also justifies why geodesics in a 
thick braneworld model with growing warp factor i.e. $f(\sigma) \sim \log(\cosh \sigma)$ are confined near $\sigma = 0$, whereas with a decaying warp factor we don't see such confinement.

\section{Summary and conclusions}

To conclude, we provide below a summary of the results obtained, in a systematic way.

\begin{itemize}
\item {We have studied geodesics for the 
warped background geometry of a thick braneworld model with a
cosmological metric on the brane and a time dependent extra dimension, in detail. 
Our choices for the scale factors (marked as set (A) and set (B) throughout this article) illustrate situations corresponding to decelerating (i.e. set (A))  and accelerating (i.e. set (B)) 
universes. The extra dimensional scale is chosen appropriately which stabilises to a constant value at large time.}
\item{The geodesic equations in such a background can be rewritten as a first 
order autonomous dynamical system. 
Analytical insights as well as some specific solutions can therefore be
obtained. In particular, the analysis of these equations as a dynamical system shows the 
role of the warp factor (growing/decaying) in controlling the nature of 
the solutions (oscillatory/exponential). }
\item{Apart from a few specific cases, the geodesic equations,
in general, can only be solved numerically since they 
are highly coupled differential equations. However, we are able to demonstrate some qualitative features of the dynamical system through numerical investigations. 
Detailed features of the null and timelike geodesics are brought out for 
different cases with growing and decaying warp factors. 
It may be noted that
a numerical estimation is always possible for any other 
combination of scale factors/warp factor.} 
\item {We have tried to separate out the individual 
effects of the warp factor 
and the extra dimensional scale factor. It is found that the emergence of an  oscillatory 
behaviour in timelike geodesics (absent in the case of null geodesics) 
is essentially due to the presence of a growing warp factor. It seems that 
massive particles are constrained to be confined near the brane in this 
particular scenario (growing warp factor), 
while massless particles like photons or gravitons can indeed
access the bulk. This observation is important in the context of localisation 
of fields on the brane. The behaviour of localisation of test particles
in a geometry with a growing warp factor is reminiscent of fermion localisation
with a growing warp factor. Even in the presence of a decaying warp factor, particles can also be localised though very special initial conditions are required. The effect of a dynamical (time dependent) extra dimension on the 
trajectories is found to be largely quantitative.}
\item {We have clearly demonstrated that the analysis of geodesic motion
obtained through a dynamical systems analysis matches quite well with
the numerical computations.}
\item {We have briefly examined a confinement condition obtained in \cite{seahra} for our cases here. We also provide our comments on aspects of this condition and show that the issue of confinement puts important constraints on possible dynamic behaviour of the extra dimension.}
\end{itemize}

We mention that it would be useful to investigate geodesic
deviation and the kinematics of 
geodesic flows in a bulk geometry with a thick brane.
In the latter case, we need to solve the geodesic and Raychaudhuri 
equations simultaneously, in order to obtain the evolution and
initial condition dependencies of the 
kinematical quantities: expansion, rotation and shear (or ESR). 
We believe that such 
work may improve our understanding about the 
nature of warped extra dimensions. We hope to report on related work
on the above-mentioned aspects, at a later stage.

\section*{Acknowledgments}
\noindent SG thanks Indian Institute of Technology (IIT), Kharagpur, India for financial support and Centre for Theoretical Studies, IIT Kharagpur, India for allowing him to use its research facilities. The authors (SK and HN) sincerely thank the Department of Science and Technology (DST), Government of India for financial support (grant number: SR/S2/HEP-10/2005). We also thank P. S. Dutta for some useful discussions.

\end{document}